\documentclass[submission,copyright,creativecommons]{eptcs}
\providecommand{\event}{CompMod 2011} 
\usepackage{array}
\usepackage{graphicx}
\usepackage{amsfonts}
\usepackage{amsmath}
\usepackage{multirow}
\usepackage{listings}

\usepackage[english]{babel}
\usepackage[tabtopcap,TABBOTCAP]{subfigure}

\title{Multiple verification in computational modeling\\ of bone pathologies}
\author{Pietro Li\`o
\institute{Computer Laboratory\\
University of Cambridge\\
Cambridge, UK}
\email{pl219@cam.ac.uk}
\and
Emanuela Merelli \qquad\qquad Nicola Paoletti
\institute{School of Science and Technology\\ Universit\`a di Camerino\\
Camerino, Italy}
\email{\quad name.surname@unicam.it}
}
\def\titlerunning{Verification instances for bone pathologies}
\def\authorrunning{Li\`o, Merelli \& Paoletti}
\begin{document}
\maketitle

\begin{abstract} 
We introduce a model checking approach to diagnose the emerging of bone pathologies. The implementation of a new model of bone remodeling in PRISM has led to an interesting characterization of osteoporosis as a defective bone remodeling dynamics with respect to other bone pathologies.
Our approach allows to derive three types of model checking-based diagnostic estimators. The first diagnostic measure focuses on the level of bone mineral density, which is currently used in medical practice. In addition, we have introduced a novel diagnostic estimator which uses the full patient clinical record, here simulated using the modeling framework. This estimator detects rapid (months) negative changes in bone mineral density. Independently of the actual bone mineral density, when the decrease occurs rapidly it is important to alarm the patient and monitor him/her more closely to detect insurgence of other bone co-morbidities. A third estimator takes into account the variance of the bone density, which could address the investigation of metabolic syndromes, diabetes and cancer. Our implementation could make use of different logical combinations of these statistical estimators and could incorporate other biomarkers for other systemic co-morbidities (for example diabetes and thalassemia).
We are delighted to report that the combination of stochastic modeling with formal methods motivate new diagnostic framework for complex pathologies. In particular our approach takes into consideration important properties of biosystems such as multiscale and self-adaptiveness. The multi-diagnosis could be further expanded, inching towards the complexity of human diseases. Finally, we briefly introduce self-adaptiveness in formal methods which is a key property in the regulative mechanisms of biological systems and well known in other mathematical and engineering areas.
\end{abstract}

\section{Introduction}

To our knowledge in the state of the art of formal methods there is no significant results for a multiscale verification, self-adaptiveness and control. Since these properties are at the core of biological processes and they are considered in modern systems biology and translational medicine, here we briefly discuss the potential of developing model checking for translational medicine, by incorporating self-adaptiveness and multi-level properties. 
In engineering, physics, meteorology, medicine, biology, social science and computer science, multiscale modeling is the field of solving physical problems which have important features at multiple scales, particularly multiple spatial and(or) temporal scales.  The self-adaptiveness is a feature that allows a system to verify properties, predict a new behavior, then to control the system, it changes the current behavior by adapting to a new situation.

Throughout life, the skeleton is continuously renewed by bone remodeling, a process which serves the purpose of repairing damaged bone and adapting the skeleton to changes in physical load. Therefore, the bone is a complex, multiscale process in which genetic mutations manifest themselves as functional changes at the cellular and tissue scale. The multiscale nature of bone requires mathematical modeling approaches that can handle multiple intracellular and extracellular factors acting on different time and space scales. Previous work has focused on a Shape Calculus \cite{shape1,shape2} approach to bone remodeling \cite{lio2011}. In this paper new models provide a way to integrate both discrete and continuous behavior that are used to represent individual cells (as agents) and concentration or density fields (as environment), respectively. Each discrete cell can also be equipped with sub-models that drive cell behavior in response to micro-environmental cues. Moreover, the individual cells (i.e. agents) can interact with one another to form and act as an integrated tissue.

\subsection{Bone pathologies altering the bone remodeling process}
\label{basic}
There are two main types of bone tissue: a compact, hard tissue which forms the outer shell of the bones and is organized in concentric layers; a second type, termed cancellous or spongy is located beneath the compact bone and consists of a meshwork of bony bars (trabeculae) with many interconnecting spaces containing bone marrow which has hematopoietic activity. Bone mass increases with growth in the first decades of life, and around the age of 30 years the peak bone mass is reached. Thereafter, as a result of mechanisms involving bone remodeling, very often a net bone loss is seen. 

In order to maintain the mechanical properties of the bones, old bone is continuously replaced by new tissue \cite{manolagas2010}. However, pathological conditions can alter the equilibrium between bone resorption and bone formation; osteoporosis, and often osteomyelitis and bone cancer, are examples of pathologies with negative remodeling: the resorption process prevails on the formation one and this reduces bone density, so increasing the risk of spontaneous fractures and delaying the recovery. 

At the basis of the bone remodeling process there is the activity of populations of cells, namely {\it osteoclasts} and \textit{osteoblasts} organized in Basic Multi-cellular Units (BMUs). Osteoblasts (the ``fillers''), which derive from pre-osteoblasts in the blood, follow osteoclasts (the ``diggers'') in a highly coordinated manner indicating that a coupled regulative mechanism must exist. Most of the osteoblasts will become \textit{osteocytes}, another type of bone cell forming a network embedded in the bone matrix and able to mechanosensing micro-fractures and to send signals to the BMU cells. This process is highly dynamic and each BMU has a finite lifetime, so new units are continuously forming as old units are finishing \cite{karsenty2010}. In the normal bone, the number of BMUs, the bone resorption rate, and the bone formation rate are all relatively constant depending on hormones, vascularization and genetics \cite{raggatt2010}. In cortical BMUs, osteoclasts excavate cylindrical tunnels in the predominant loading direction of the bone. They are followed by osteoblasts, filling the tunnel, creating secondary osteons of renewed tissue. 

The \textit{RANK/RANKL/OPG} signaling pathway plays an important role in bone metabolism. RANK is a protein expressed by osteoclasts and acts as a receptor for RANKL, a protein produced by osteoblasts. RANK/RANKL binding induces osteoclast differentiation, proliferation and activation. Osteoprotegerin (OPG) is a decoy receptor for RANKL. It is expressed by mature osteoblasts and it binds with RANKL, and preventing it from binding to RANK. In this way, OPG inhibits the production of osteoclasts, thus protecting bone from excessive resorption. The relative concentration of RANKL in bone is a major determinant of bone mass and strength \cite{hanada2010}. In particular, recent works show that the levels of OPG and RANKL are inversely related to bone density, and may lead to the development of osteoporosis~\cite{jabbar2011}.

\begin{figure}
\centering
\includegraphics[width=8cm, angle=0]{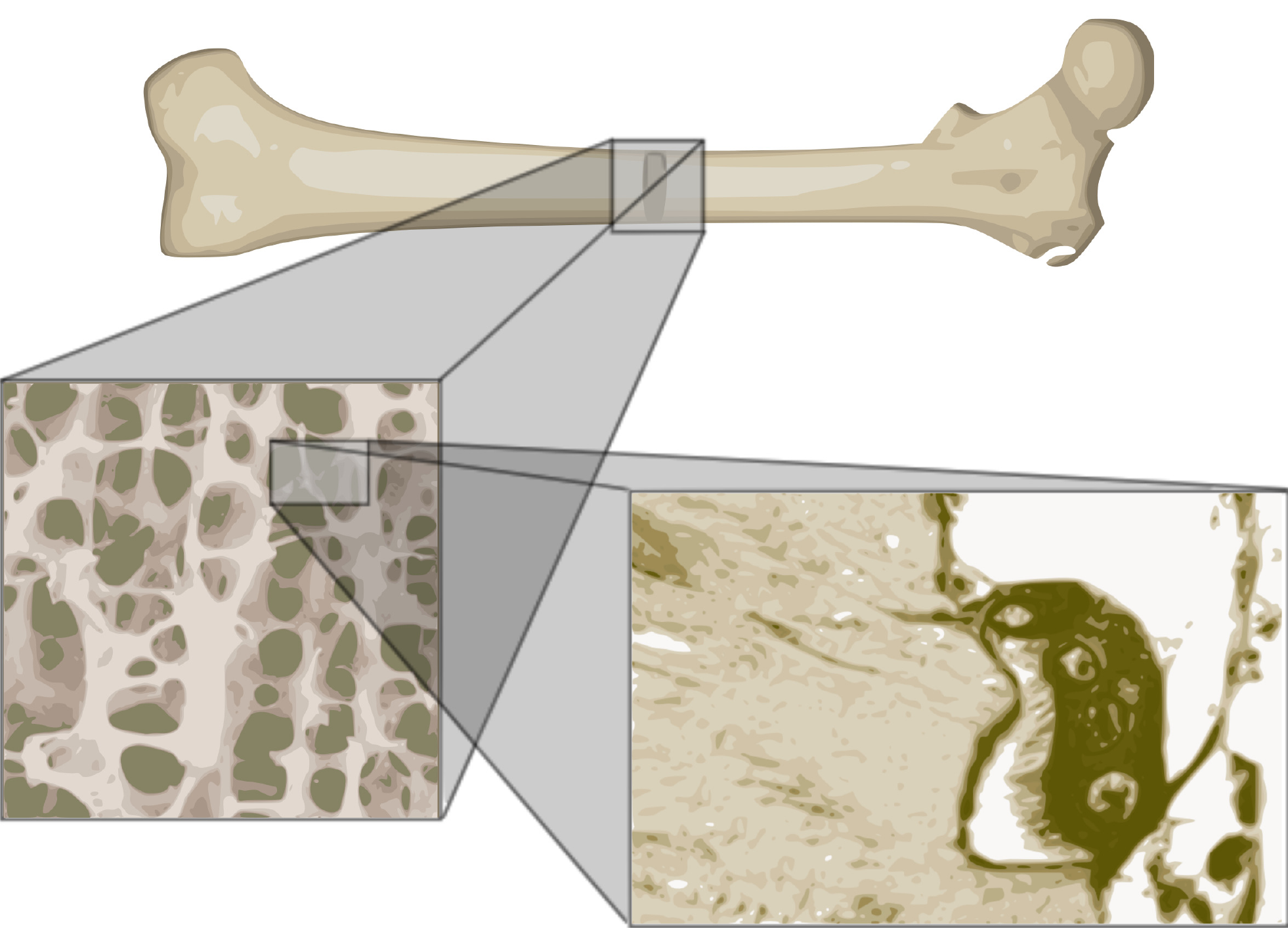}
\caption{Multiscale representation of human femur}
\label{femur}
\end{figure}

\begin{figure}[ht]
\centering
\includegraphics[width=9cm, angle=0]{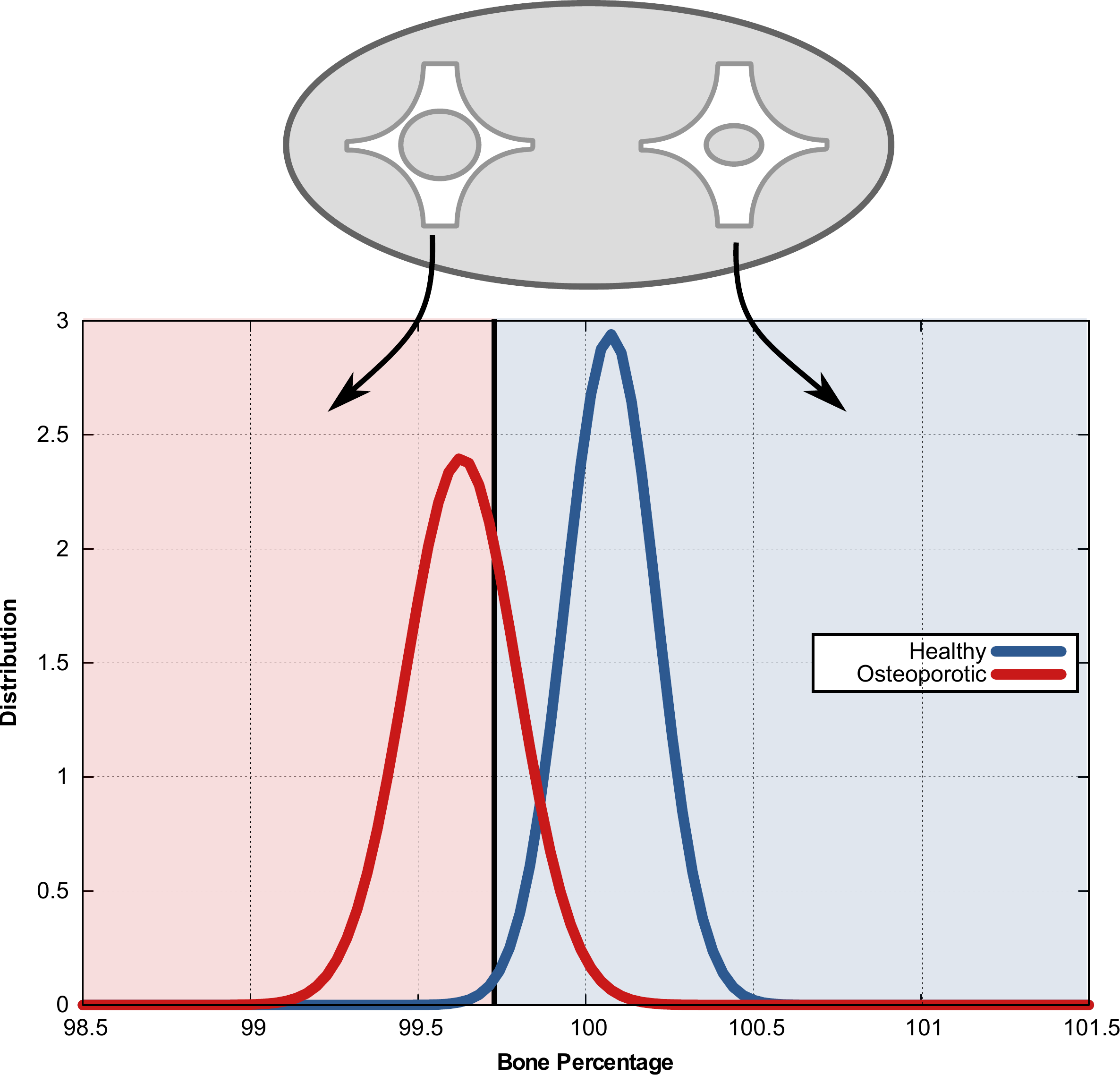}
\caption{Bone density distribution in healthy and osteoporotic patients (from literature). Our model considers that healthy and osteoporotic patients have slightly different parameter space.}
\label{comparison}
\end{figure}

Our previous work~\cite{paoletti2011} shows how small changes in RANKL for very long periods lead to disease conditions, especially when aging factors are involved. In particular, we have simulated the bone remodeling process, defining two subsets in parameter space:
\begin{itemize}
\item \textbf{healthy configuration}, where RANKL production and cellular activity is normal, and an
\item \textbf{osteoporotic configuration}, with an overproduction of RANKL and a reduced cellular activity.
\end{itemize}
In this setting, the overproduction of RANKL is indicative of an inflammation or of hormonal imbalance, and the lower cellular activity is typical for older people. 
In the osteoporotic case, higher values of RANKL promote a higher osteoclastic activity, leading to a higher resorption. Moreover, the reduced osteoblastic activity is not sufficient to completely repair the consumed part of bone. This causes a bone density net loss, weakening trabeculae and consequently causing micro-fractures in osteoporotic patients. Note that in a young patient, an overproduction of RANKL doesn't determine a disease situation. In fact, an unexpected high resorption activity can be easily balanced by recruiting more osteoblasts, which is not possible in an old patient. On the other hand, with normal RANKL concentrations, fewer osteoclasts are balanced by even fewer osteoblasts (see Figure \ref{comparison} for a comparison of bone mineral density in healthy and osteoporotic conditions).

The remainder of the paper is organized as follows. In Section 2 we provide a PRISM model for bone remodeling. In Section 3 we discuss the self-adaptiveness in biological systems, and the potentialities for translational medicine modeling. Finally we draw some conclusions and sketch possible future work.
\section{A PRISM Model for Bone Remodeling}
Here we will formally assess the differences in bone density by evaluating the dynamics of the bone remodeling process being in a certain state. This kind of data can be elaborated by means of automatic verification procedures~\cite{kwiatkowska2002prism}. We believe that this kind of quantitative, formal and automated analysis may represent a step ahead in the understanding of the modeling efforts of osteoporosis development, by  shifting the attention from an informative, but empirical, analysis of the graphs produced by simulations towards more precise quantitative interpretations.

Many models describing the behavior of dynamical systems, like distributed and concurrent computational systems, have been equipped with a formal semantics that precisely describes the possible evolution of the system. Often, such a semantics is given in terms of a Transition System that defines the possible states of the modeled systems and the modalities to move from one state to another. Recently model checking methods have been introduced in the analysis of genetic networks; here we further extend to cell networks and translational medicine. Model checking is perhaps one of the most established ``proof" techniques in symbolic reasoning. The concept of proof gets instantiated appropriately through an automatic search for a property (specified as a logical formula) to hold or not over a finite set of states.
The outcome of model checking is either an affirmative answer or a counter example, e.g. a set of states representing possible evolutions of the system that do not fulfill the formula.
\subsection{Probabilistic Model Checking and PRISM}
\textit{Probabilistic model checking} is a variant of classical model checking techniques and it is typically used when the system to model exhibits a random or probabilistic behavior. Its output is not simply an affirmative or a negative answer, but it allows to verify quantitative properties over the model: instead of \textit{``will the system eventually reach a particular state?''}, we can verify properties like \textit{``what is the probability that the system eventually reach a particular state?''}.

Probabilistic verification techniques have proved to be particularly suitable in the analysis of biological systems, which are intrinsically random and stochastic. Living cells are complex mixtures of a variety of complex molecules that are constantly undergoing reactions with one another, and such reactions typically have an exponential distribution associated~\cite{gillespie1977exact}. Furthermore, stochastic fluctuations play a key role in biological processes, both at the molecular level, where low intracellular copy numbers of molecules can fundamentally limit the precision of gene regulation e.g. in gene expression dynamics; and at the macroscopic level, e.g. driving the cells' phenotypic state. In the process of bone remodeling, the statistical fluctuations in RANKL concentrations in the blood produce changes in the chemotaxis, i.e. the process by which cells move toward attractant molecules, of osteoclasts and osteoblasts. This may affect for example the cell differentiation, number and arrival time, and consequently the whole remodeling process, as explained in Section~\ref{basic}.

In probabilistic model checking models are enriched with quantitative information, i.e probabilities or stochastic rates, representing the ``propensity'' with which a certain transition occurs. Such quantities affect the stochastic dynamics as expected, since dynamics is ruled by the law of mass action, as standard (according to this law, the strength of a reaction, its probability to occur quickly say, depends on the associated rates and the amount of entities ready to participate into the reaction). Typical modeling techniques are \textit{Markov chains} and \textit{Markov decision processes}. Logical formulas are equipped with quantities as well, and their execution returns a probability or a reward value instead of a boolean value.

In this work we have adopted the open-source PRISM probabilistic model checker~\cite{kwiatkowska2002prism}, one of the reference model checkers for the analysis of stochastic and probabilistic systems. PRISM models are specified in a formal language that describes the entities present in the model as \textit{modules}. Each module is characterized by a set of \textit{state variables} and a list of \textit{guarded and stochastic commands}. Then, the tool builds the concrete model from the PRISM specification. Currently the following models are supported: discrete-time Markov chains, continuous-time Markov chains, Markov decision processes, probabilistic automata, and probabilistic timed automata.

The PRISM model checker have been successfully employed in a several number of biological case studies, especially for modeling biochemical pathways~\cite{kwiatkowska2008using,calder2006analysis,barbuti2005probabilistic,pronkastochastic}. However, our framework is not determined by the features of a specific tool, and other similar platforms could have been chosen.

\subsection{Definition of the model}
In our settings, we use \textit{Continuous Time Markov Chains (CTMC)}, and the involved entities, or \textit{modules} are {\em osteoclasts} and {\em osteoblasts}. Each state typically consists of the values of the local and global variables of the model, which keep trace of the amount and variations of involved entities. Listing~\ref{lst:code} shows the PRISM code for {\em bone remodeling}. We employ a \textit{population-based} approach: a state of the system is characterized by the number of osteoclasts (variable {\tt Oc}) and osteoblasts (variable {\tt Ob}) and by their status, that determines whether they are precursors or mature cells (lines 8 and 17). In order to cope with the state-explosion problem, we do not consider the actual concentrations of osteoclasts ($\approx$ 10 cells/BMU) and osteoblasts ($\approx$ 1000 cells/BMU). We rather consider a sub-volume of the BMU, characterized by smaller ranges: $0-5$ for {\tt Oc} and $0-100$ for {\tt Ob}.

As mentioned in~\ref{basic}, a reduced cellular activity and an overproduction of RANKL could be the main causes of osteoporosis. We model these two factors as parameters, {\tt aging} and {\tt rankLrate}, by varying which we can run and analyze 
\begin{itemize}
\item a healthy configuration: {\tt aging = 1, rankLrate = 0.1}, and 
\item a sick configuration: {\tt aging = 2, rankLrate = 0.2}.
\end{itemize}
Then, several possible actions, corresponding to the participation to a reaction, are defined. Each consists of a name, a guard, a rate and an effect. In the multicellular settings of our model, actions can trigger an internal state change (e.g. cell differentiation), or can synchronize with other actions in order to express cellular signaling mechanisms (e.g. pre-osteoclasts' proliferation induced by pre-osteoblasts). Resorption and formation are also implemented as actions, and the quantity of bone consumed or formed is computed by using a \textit{transition reward} on those actions. On the one hand, the reward structure allows us to express real-valued bone densities, with respect to state variables which can only assume discrete values. On the other hand, we keep a lightweight model, since rewards do not increment the state space differently from variables.

In the initial state, osteoclasts and osteoblasts are precursors (\texttt{pc:bool init true; pb:bool \\init true;}). At this point, pre-osteoblasts proliferate linearly (line 18) and after they have reached a concentration threshold, they start producing RANKL, by performing an action {\tt [rankl]} (line 19). RANKL triggers pre-osteoclasts' proliferation by synchronizing on {\tt [rankl]} (line 9). \\Then, pre-osteoclasts can become mature osteoclasts (line 10) and start consuming bone with the \\{\tt [resorb]} action (line 11). After each {\tt [resorb]} action, an osteoclast dies and this induces the maturation of pre-osteoblasts (line 20). Mature osteoblasts mineralize bone (line 21) and similarly to osteoclasts, each {\tt [form]} action causes the death of an osteoblast. Once the concentration of osteoclasts and osteoclasts is null, they turn back into precursors and the remodeling cycle starts again.
\lstset{ %
basicstyle=\scriptsize,       
showspaces=false,               
numbers=left,                   
numberstyle=\scriptsize,      
showstringspaces=false,         
showtabs=false,                 
frame=single,                   
captionpos=b,                   
breaklines=true,                
breakatwhitespace=false,        
caption=Core PRISM model for bone remodeling,
label=lst:code
}

\begin{lstlisting}
const double aging;
const double rankLrate;
const double formRate = 0.03/aging;
const double resorbRate = 5*rankLrate/aging;

module osteoclasts
Oc:[0..5] init 0;
pc: bool init true;
[rankl] pc=true & Oc<5 -> Oc+0.1:(Oc'=Oc+1);
[] pc=true & Oc>1 -> 1:(pc'=false);
[resorb] pc=false & Oc>0 -> resorbRate*pow(Oc,2):(Oc'=Oc-1);
[] pc=false & Oc=0 -> 1:(pc'=true);
endmodule

module osteoblasts
Ob:[0..100] init 1;
pb: bool init true;
[] Ob>0 & Ob<100 & pb=true -> pow(Ob,0.5):(Ob'=Ob+1);
[rankl] pb=true & Ob>50-> rankLrate*Ob:true;
[resorb] pb=true -> 1:(pb'=false);
[form] Ob>0 & pb=false -> formRate*Ob:(Ob'=Ob-1);
[] pb=false & Ob=0 -> 1: (pb'=true) & (Ob'=1);
endmodule

rewards "boneResorbed"
[resorb] true:resorbRate;
endrewards

rewards "boneFormed"
[form] true:formRate;
endrewards
\end{lstlisting}
\subsection{Definition of the properties}
In this work we are interested in analyzing and verifying some crucial properties of the model, and compare the results between healthy and pathological scenarios over a time lapse of four years, which is enough to assess the presence of bone diseases. We do not focus on cellular properties (e.g. the probability that osteoclasts are mature at a given time), but rather we aim to verify the \textit{emerging properties} of the bone tissue (e.g. the rapidity of negative remodeling) which are more relevant from a clinical perspective. 
In particular, we introduce a statistical estimator based on bone mineral density which reflect the tissue level. A second estimator detects rapid (months) decreases. This estimator could derive from different cofactors (morbidity), for example diabetes or cancer. It is therefore more linked to disruptions at the level of cell-cell interaction and abundances. A third estimator takes into account the variance of the bone density. The variability of the bone density could address the investigation of metabolic syndrome, diabetes and cancer.
The logical properties to verify have been formulated in \textit{CSL (Continuous Stochastic Logic)}~\cite{AS+00}, and are:
\begin{itemize}
\item The cumulative values of bone formation and resorption. 
\begin{align*}
f_+(t): & \ R \lbrace``boneFormed''\rbrace=?[C\leq t],\\
f_-(t): & \ R \lbrace``boneResorbed''\rbrace=?[C\leq t], \quad t = 0,10,\ldots,1460.
\end{align*}
The results of the verification for the healthy and the osteoporotic case are displayed in Figure~\ref{fig:prism:formres}. Note that in normal conditions the disruptive and constructive activities grow with the same magnitude, while in disease conditions resorption increase more quickly than formation.
\item The progress of bone density, intended as the difference between bone formation and bone resorption. $$f_{BD}(t): f_+(t) - f_-(t), \quad t = 0,10,\ldots,1460.$$
Figure~\ref{fig:prism:bonemass} shows how in the healthy case the bone mass tends to a steady level, differently from the osteoporotic case where bone mass decreases linearly.
\item We are interested to verify not only the expected values of bone mass, but also the variance of bone density with respect to the values from each state, at $t=365$ (1 year), $t=730$ (2 years), $t=1095$ (3 years), and $t=1460$ (4 years). Since we are working with rewards, we cannot express this property with a $P$-formula, but we are able to compute the states satisfying this property. We make use of PRISM filters, a particular kind of formulas capable to compute values simultaneously for several states. Filters are of the form $$filter(op,prop,states),$$ where $op$ is the operator characterizing the type of filter (max, min, avg, range, count, \ldots); $prop$ is the PRISM property to verify; and $states$ is the predicate identifying the set of states over which to apply the filter (if $true$ it can be omitted). In order to compute all the density values from all the states, we make use of the $print$ filter, a particular operator that prints in the PRISM log each state and the value from $prop$.
$$filter(print, f_{BD}(t)< \mathbf{d}), \quad t=365,730,1095,1460 \quad \ \mathbf{d}=-5,-4.5,\ldots,4.5,5.$$

The curves in Fig.~\ref{fig:prism:distribution} are the normal distributions with average and variance computed from the values outputted by PRISM. The results clearly show that in the healthy case the bone density is mainly distributed in the interval $[0,2]$. While in the osteoporotic case, the negative bone density gets worse and sharper as the years go by.
\item Finally, we consider the rapidity of negative changes in bone mass as a valuable indicator in assessing osteoporosis. In particular we want to verify if the value of bone density decreases of a quantity $\mathbf{k}$ between time $t$ and time $t + \Delta t$. The formula is the following.
$$f_{BD}(t+\Delta t) + \mathbf{k} < f_{BD}(t), \quad t=0,50,\ldots,1450, \quad \mathbf{k}=0.25,0.50.$$
Clearly, true results predict alarming rapid changes, which can be monitored in order to prevent osteoporosis. By increasing the threshold $\mathbf{k}$, we relax the condition so that we obtain a minor number of true results. On the other hand, the fact that the formula is satisfied with high values of $\mathbf{k}$ indicates a severe bone loss. Table~\ref{prism_rapidity1} shows the results of this analysis, comparing the healthy and the osteoporotic case and evaluating rapid negative changes in a time interval $\Delta t$ of 100 days.

Similarly, a measure able to predict also rapid positive changes would be useful in medical applications, for those pathologies characterized by a positive bone metabolism (e.g. bone metastasis). This measure has been formulated as the difference quotient of bone mineral density:
$$\frac{f_{BD}(t+\Delta t) -  f_{BD}(t)}{\Delta t}, \quad t=0,50,\ldots,1450.$$
We have verified this on our model by setting $\Delta t = 100$. Results in Fig.~\ref{prism_rapidity2} illustrate that in the healthy case only an initial rapid bone growth is registered (at $t=100$, corresponding to the interval $[100,200]$ days); thereafter the change rate reaches a stationary level close to zero. While in the osteoporotic case, alarming negative changes are reported around the interval $[0,300]$ days. Then, the curve tends to a steady level, even if the rate change is negative, confirming the net bone loss of the osteoporotic series in Fig.~\ref{fig:prism:bonemass}.

\begin{figure}
\centering
\includegraphics[width=0.95\textwidth]{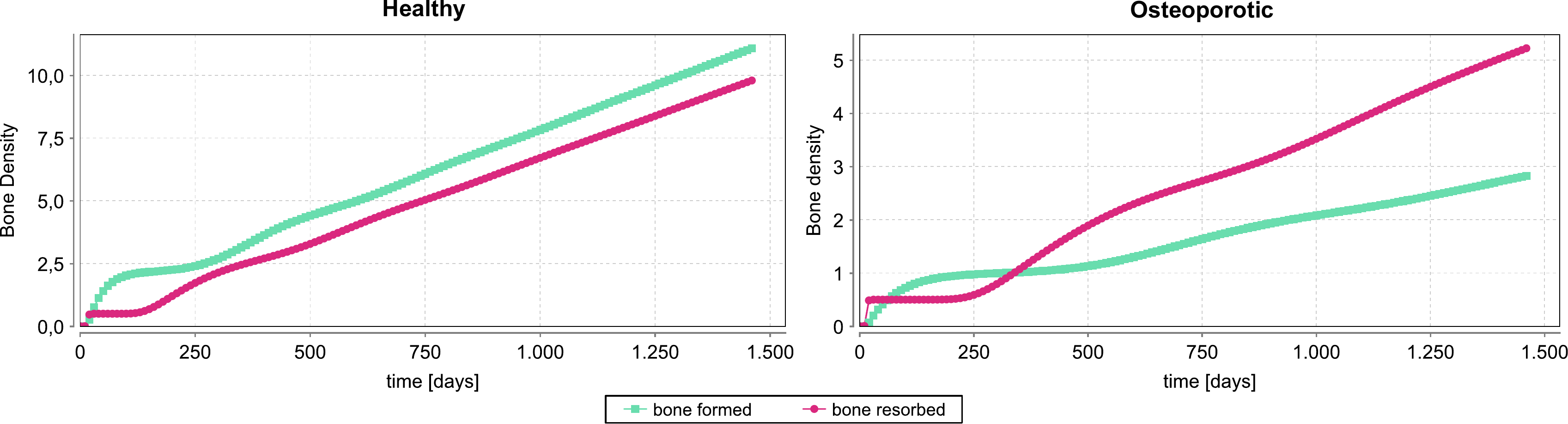}
\caption{Expected values of resorbed and formed bone in 4 years.}
\label{fig:prism:formres}
\end{figure}

\begin{figure}
\centering
\includegraphics[width=0.75\textwidth]{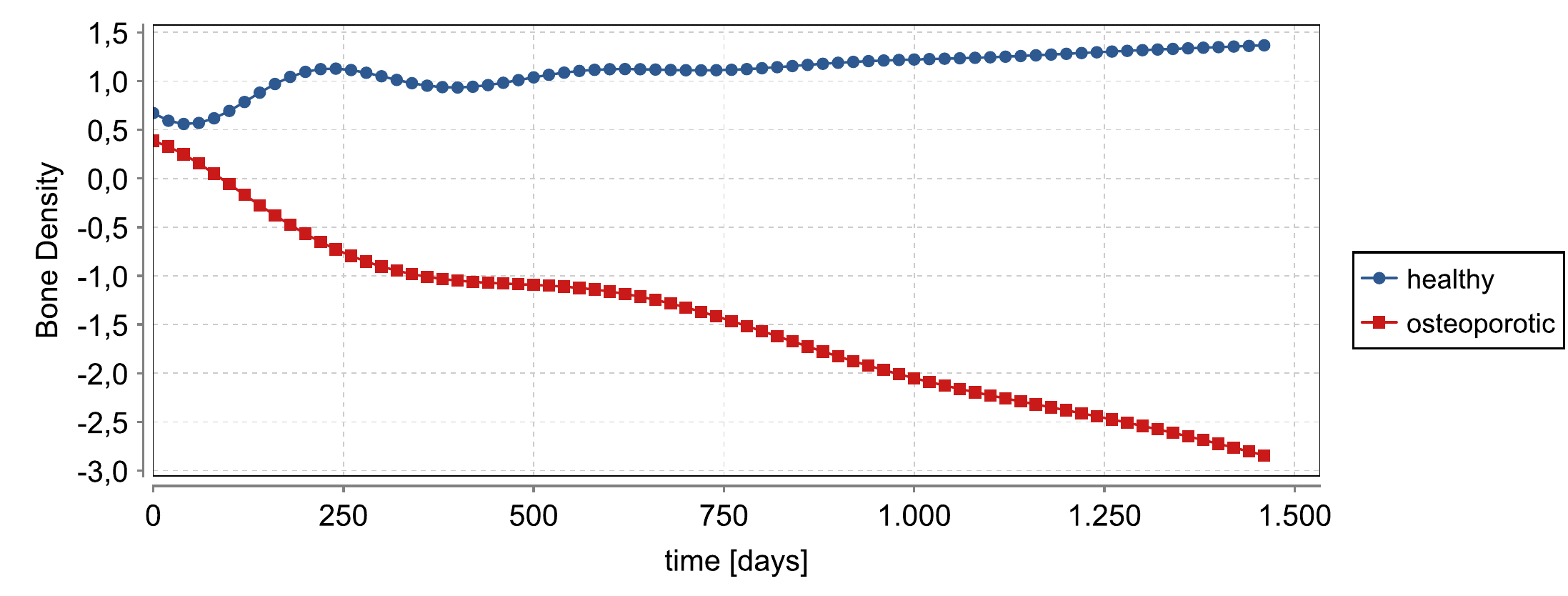}
\caption{Variation in bone density during 4 years.}
\label{fig:prism:bonemass}
\end{figure}
\begin{figure}
\centering
\includegraphics[width=0.7\textwidth]{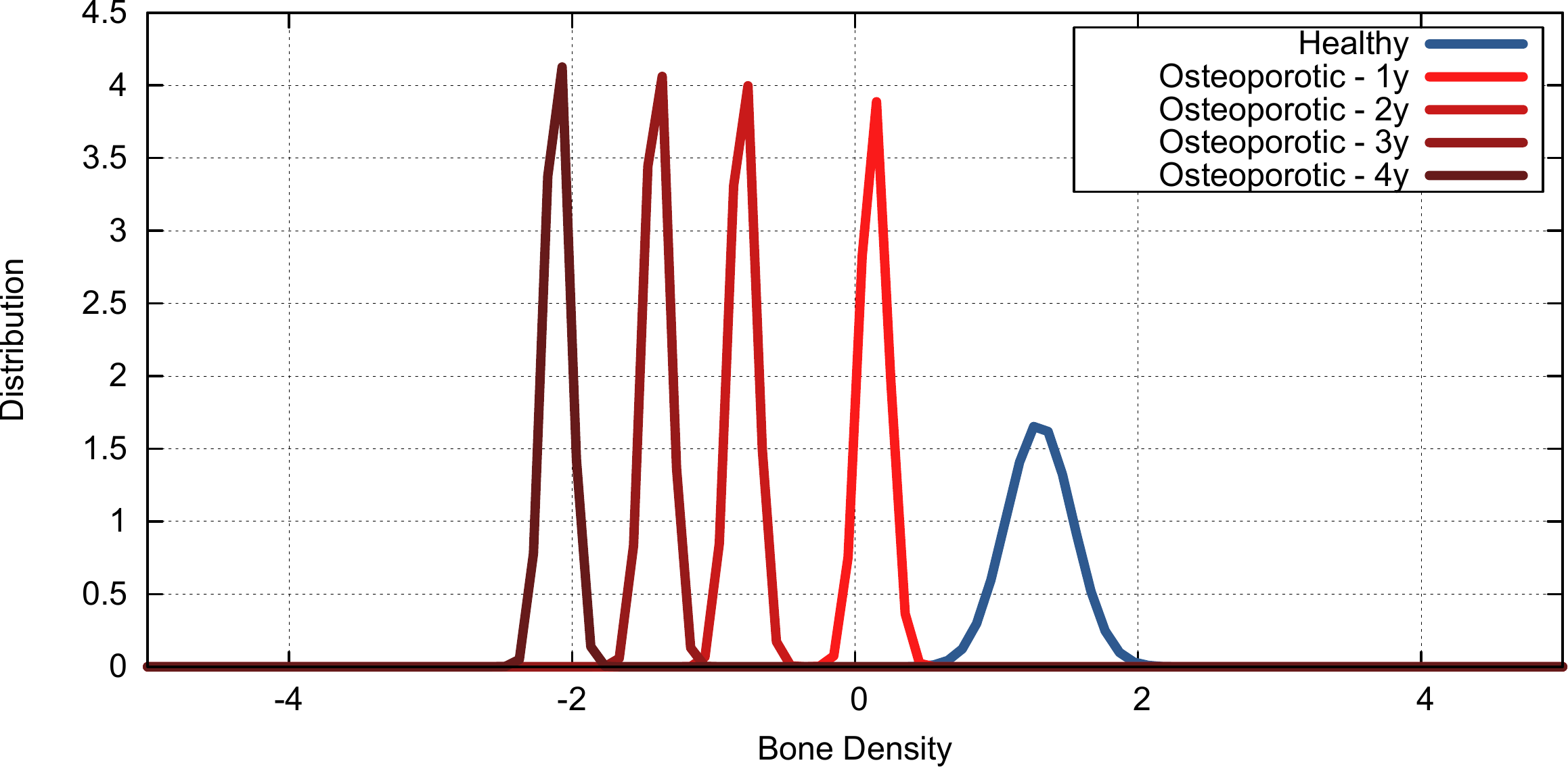}
\caption{Bone density distributions in the years.}
\label{fig:prism:distribution}
\end{figure}

\begin{table}
\centering
\begin{small}
\begin{tabular}{|c|c|c|c|c|}
\hline
\multirow{2}{*}{\textbf{$(t,t+100)$ days}}&\multicolumn{2}{|c|}{\textbf{Healthy}}&\multicolumn{2}{|c|}{\textbf{Osteoporotic}}\\
\cline{2-5}
& \textbf{0.25} & \textbf{0.5 }& \textbf{0.25} & \textbf{0.5}\\
\hline
0 & $\mathbf{tt}$ & $ff$ & $ff$ & $ff$\\
50 & $ff$ & $ff$ & $\mathbf{tt}$ & $ff$\\ 
100 & $ff$ & $ff$ & $\mathbf{tt}$ & $\mathbf{tt}$\\ 
150 & $ff$ & $ff$ & $\mathbf{tt}$ & $\mathbf{tt}$\\ 
200 & $ff$ & $ff$ & $\mathbf{tt}$ & $ff$\\ 
250 & $ff$ & $ff$ & $\mathbf{tt}$ & $ff$\\ 
300 & $ff$ & $ff$ & $ff$ & $ff$\\ 
\ldots & \ldots & \ldots & \ldots & \ldots\\
750 & $ff$ & $ff$ & $ff$ & $ff$\\ 
800 & $ff$ & $ff$ & $\mathbf{tt}$ & $ff$\\ 
850 & $ff$ & $ff$ & $\mathbf{tt}$ & $ff$\\ 
900 & $ff$ & $ff$ & $ff$ & $ff$\\
\ldots & \ldots & \ldots & \ldots & \ldots\\
1450 & $ff$ & $ff$ & $ff$ & $ff$\\ 
\hline
\end{tabular}
\end{small}
\caption{Rapidity of negative remodeling. True values are bad (too rapid negative changes). Rows with false values in each cell have been omitted.}
\label{prism_rapidity1}
\end{table}

\begin{figure}
\centering
\includegraphics[width=0.8\textwidth]{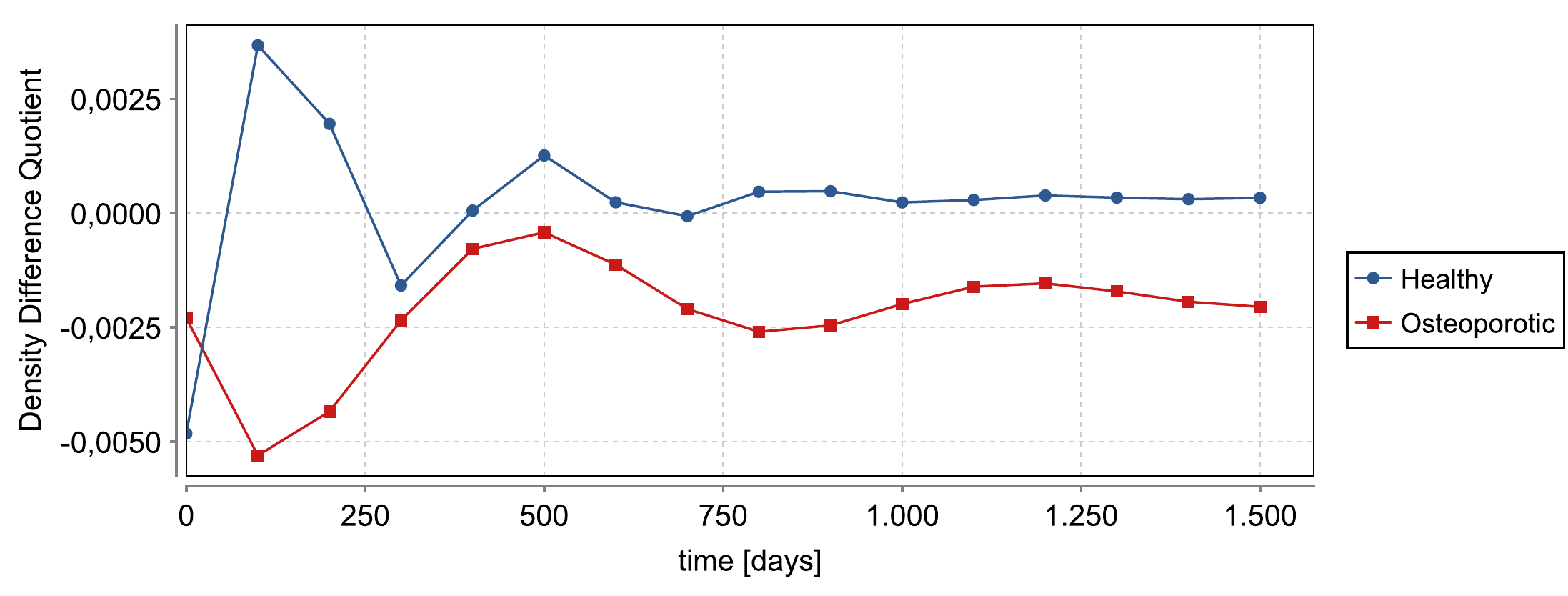}
\caption{Difference of bone mineral density. The plot shows the change rate in an interval of $\Delta t = 100$. Alarming pathological values are below -0.0025 and above 0.0025.}
\label{prism_rapidity2}
\end{figure}
\end{itemize}
\subsection{A verification-driven diagnosis}
\begin{figure}
\centering
\begin{small}
\begin{tabular}{m{1cm}m{120pt}m{120pt}}
t [days] & t=130 & t=350\\
BONE MASS \textbf{H} & \subfigure[]{\includegraphics[width=120pt, height=80pt]{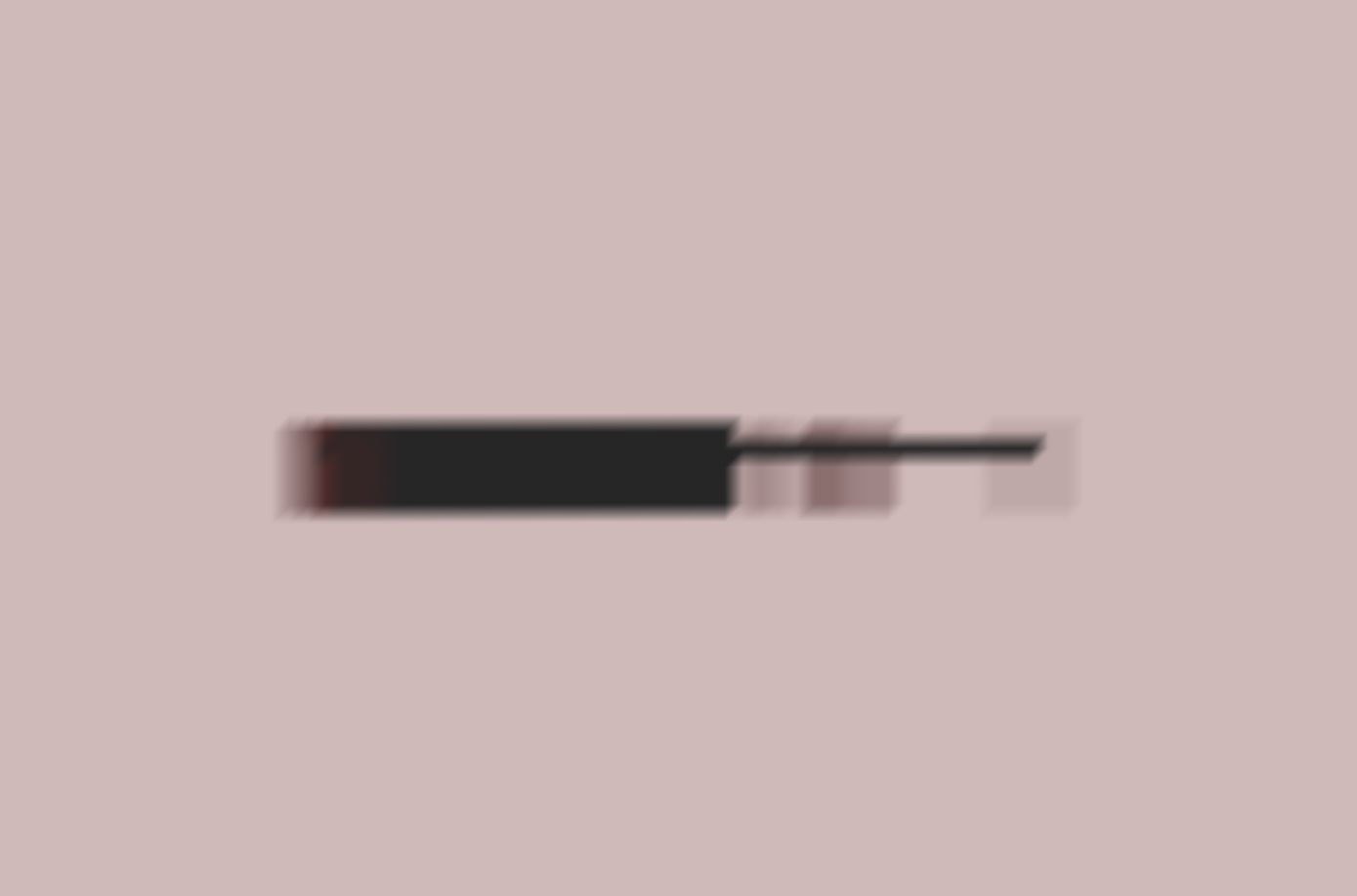}}& \subfigure[]{\includegraphics[width=120pt, height=80pt]{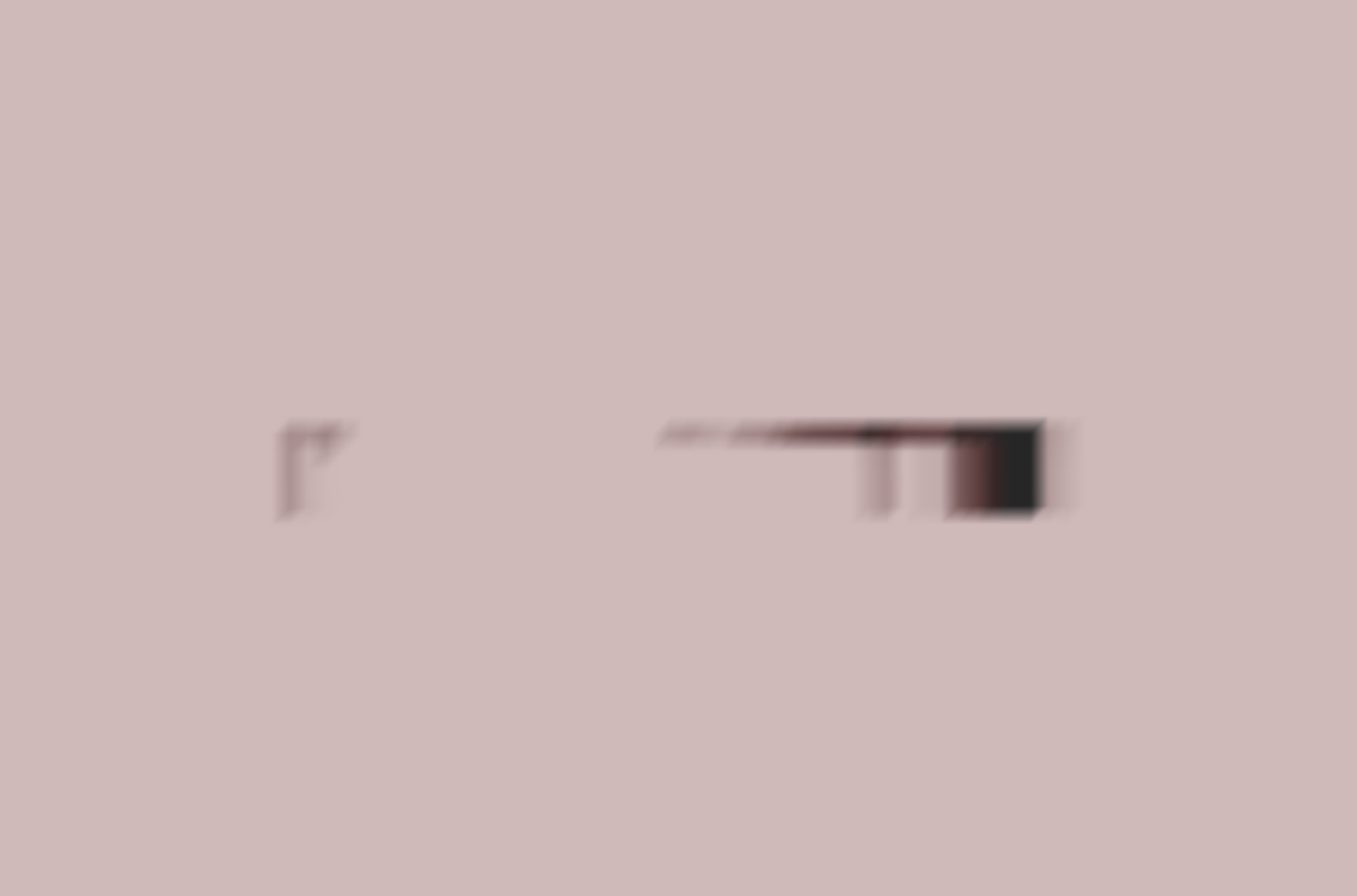}}\\
BONE MASS \textbf{O} & \subfigure[]{\includegraphics[width=120pt, height=80pt]{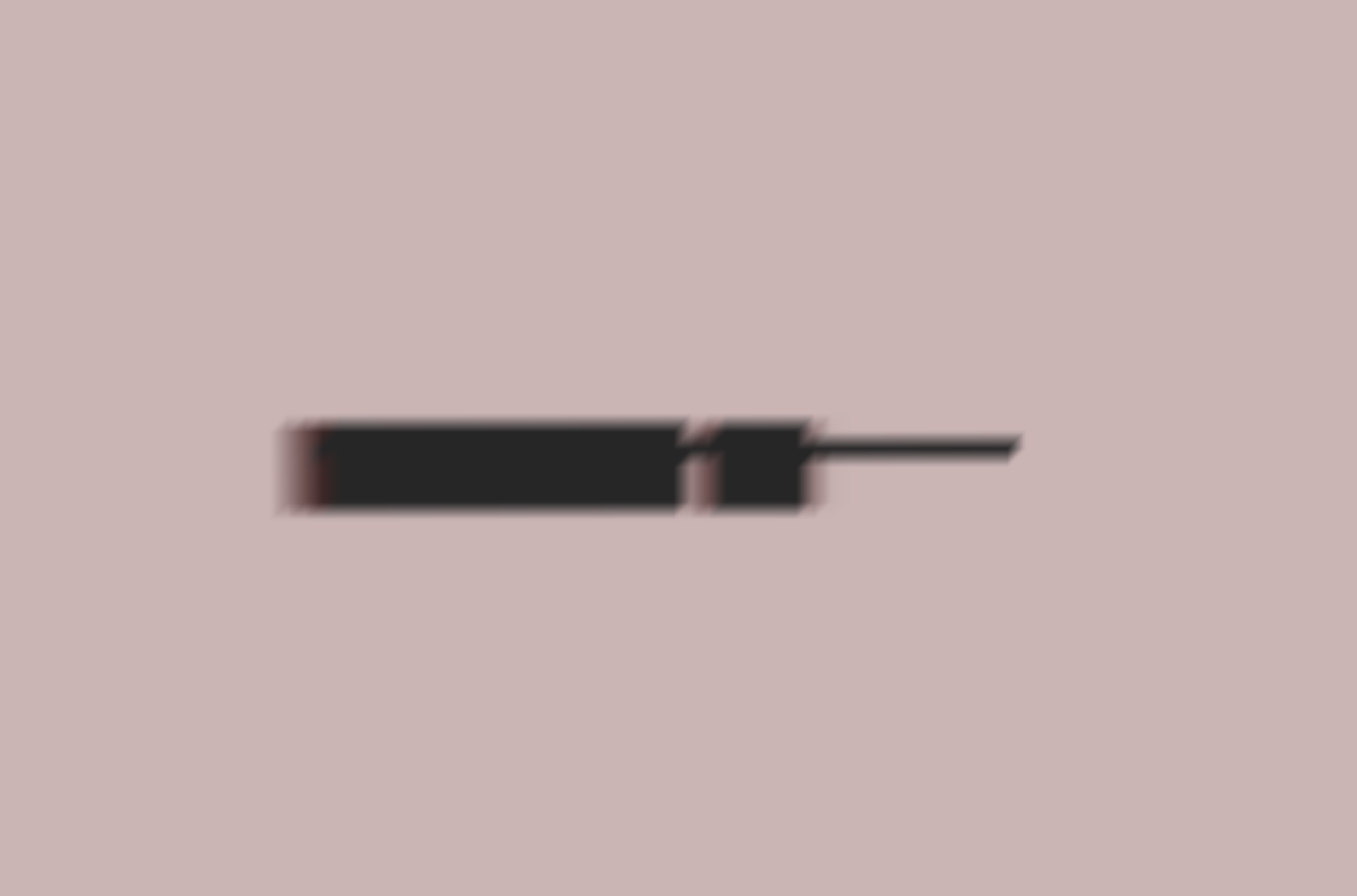}}& \subfigure[]{\includegraphics[width=120pt, height=80pt]{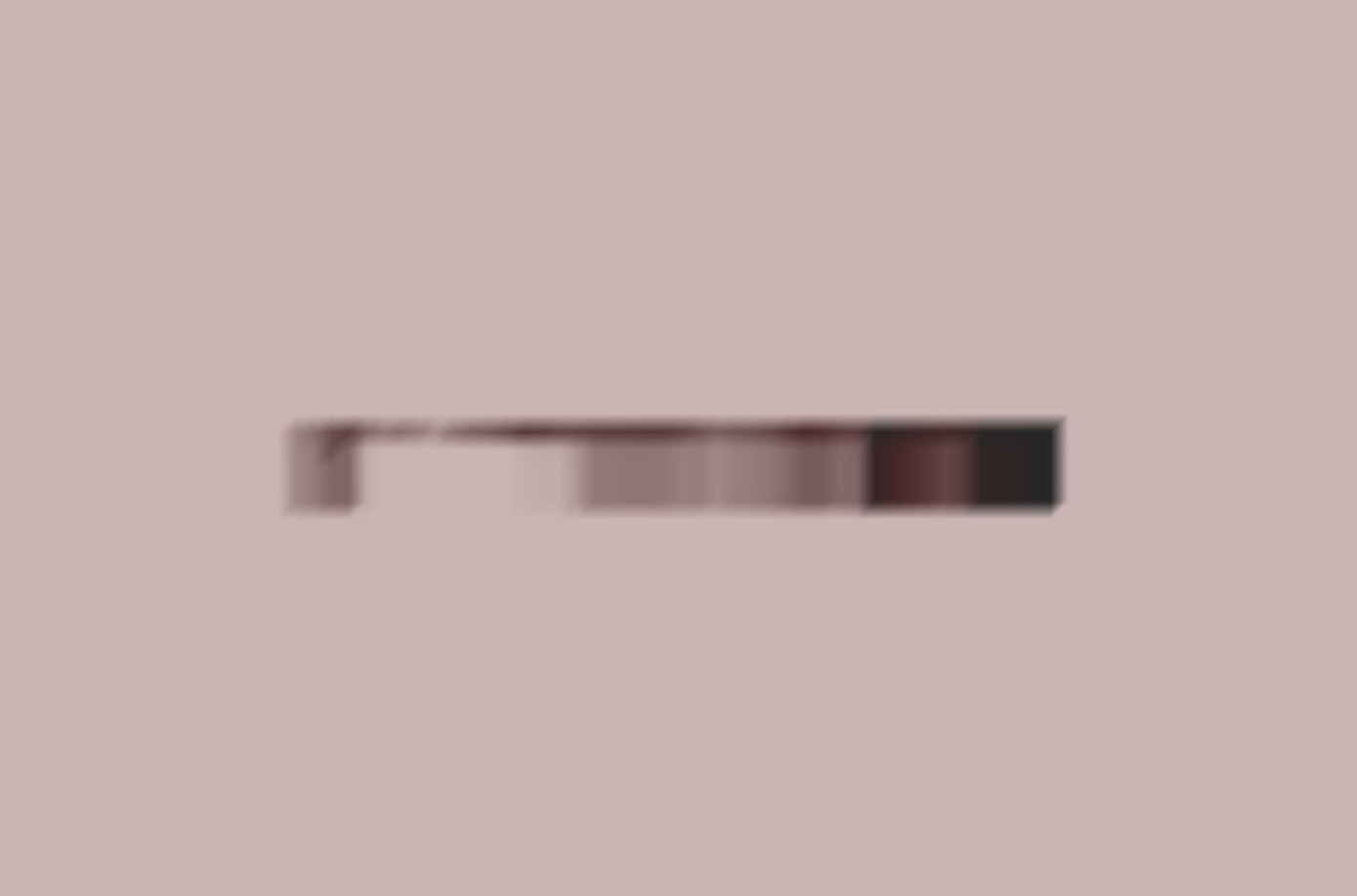}}\\
&\multicolumn{2}{c}{\includegraphics[height=10pt]{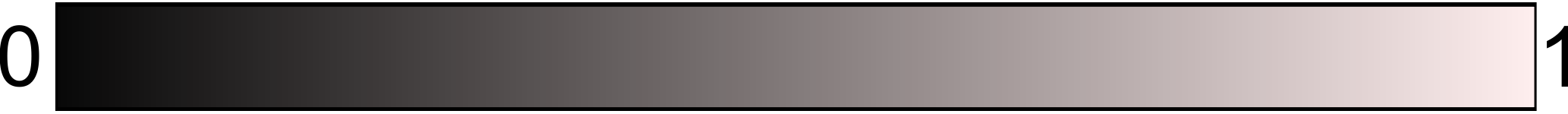}}
\end{tabular}
\caption{Bone density during the simulation of healthy (\textbf{H}) and osteoporotic (\textbf{O}) configurations for two different simulation times (130 and 350 days)}
\label{simulationScreens}
\end{small}
\end{figure}

Figure~\ref{simulationScreens} illustrates the density values of the trabeculae computed in the simulation of the two configurations, during the resorption phase ($t=130$) and in the last days of the remodeling cycle ($t=350$).
While at $t=130$ the bone density is approximately the same, at the end of the simulation differences are quite prominent, since the reduced osteoblastic activity generates several zones with a lower density.
Let assume that the simulation of the PRISM implementation of the model is
the analogue of the determination of clinical parameters during periodic
frequent medical visits.  This approach suggests that in the future the determination of the values for certain cell type and signaling molecules concentrations (perhaps measured from gene expression data) will provide means for simulating the progression of the disease and the efficacy of therapies. Therefore our work is meaningful in perspective of a new type of medicine characterized by a close coupling between clinical measures and modeling prediction.

We have defined meaningful statistical estimators: a) simple bone mineral density (amount of matter per
cubic centimeter of bones, measured as z-score, the number of standard
deviations above or below the mean for the patient's age, sex and
ethnicity; or as t-score, the number of standard deviations above or below
the mean for a healthy 30 year old adult of the same sex and ethnicity as
the patient). This is a tissue-level estimator.

b) The rate of decrease of bone mineral density during consecutive medical visits.
This estimator tells us the emergence of defects of the bone metabolism in terms of
signaling networks of RANK/RANKL and decrease of pre-osteoblast number. 
This estimator is at cell and molecular signaling levels.

c) The changes in the variance or in the fractal pattern of the signal (measured
using Hurst coefficient, for example using wavelets). If the variability of
the bone density largely increases that pattern could depend on metabolic syndrome, diabetes,
cancer. This could be mainly related to intracellular signaling
disruption. The signal could also change its fractal pattern which could be
in general related to a decrease in signaling responsiveness.
Therefore a), b) and c) tell us of different types of disruptions.
Note that there may be dependencies between a), b) and c).

Diagnosis could be different for different combinations of a), b) and c). For example:
\begin{itemize}
\item a) OR b) = ``osteoporosis or progression to osteoporosis''; 
\item a) AND b) = ``severe osteoporosis with
loss of calcium''; 
\item a) AND c) = ``severe decrease of general metabolic functions
due to important infection'';  
\item b) AND c) = ``infection (osteomielites) and/or
cancer''. 
\item d) diabetes AND b) = great risk of progressive osteoporosis;
\item e) thalassemia AND b) = great risk of osteoporosis.
\end{itemize}

Our approach is novel because it provides a complex diagnostic framework which could incorporate additional biomarkers. 
Therefore these combinations provide means for multi-disease diagnosis and multiscale causes identification.

It is noteworthy that also the shape of the trabeculae could be object of property verification. In general large part of diagnosis of pathologies depends more on visual inspection and microscopic analysis of tissues than with molecular markers. Therefore, the Shape Calculus \cite{shape1,shape2} could in principle be used to assess properties on the shape of the trabeculae.

\section{Discussion}
The computational modeling of osteoporosis is particularly challenging because the disease could span several years and decades. Although we have not addressed data fitting with individual patient data, we have used knowledge derived from literature and held regular meetings with medical scientists at the Rizzoli Hospital in Bologna for bone diseases which has a prestigious medical record in terms of number of patients, advanced technology and expertise.

In this work we raise the challenge of effectively applying formal methods in biomedical practice, showing that probabilistic model checking can be successfully employed in the diagnosis and in the prediction of complex bone pathologies like osteoporosis. The verification of quantitative properties on bone density has provided three different statistical estimators: the first one is related to bone mass; the second one monitors rapid negative (or positive) changes as a symptom of osteoporosis; the third estimator measures the variance in bone density. 

Such estimators have been evaluated over two different subsets of the parameter space of the model, corresponding to a healthy configuration and an osteoporotic configuration. The results indicate that in the first case bone homeostasis, i.e. the balance between resorption and formation, is maintained during the years and that the variance of bone density is delimited to a well-defined interval. On the other hand in the osteoporotic case, we experience a net bone loss which is constant throughout the years, also characterized by high peaks of negative change rate. Furthermore we show how more complex pathologies can be diagnosed by combining these three estimators. Besides being suitable to assess other types of diseases affecting bone remodeling, our approach represents an innovative diagnostic framework which can be of inspiration for a new type of medicine which combines clinical measures and modeling predictions.

As future work, a desired direction would be reverse engineering from patient datasets, which in turn will allow to fine tuning the diagnosis. In addition, we aim at using parametric probabilistic model repair techniques~\cite{bartocci2011model} for our model. The problem of model repair is, roughly speaking, ``given a probabilistic model $M$ and a formula not satisfied by $M$, finding a $M'$ that satisfy the formula and such that the cost of modifying $M$ to obtain $M'$ is minimized''. Consider our model of bone remodeling, and a formula for checking if the system will reach a required bone density. Suppose that the formula is not satisfied, as in an osteoporotic patient. The idea is that parametric model repair can automatically suggest the minimal changes to adopt in the parameters (e.g. RANKL or cellular activity) such that the system can reach that desirable bone density. We believe that this could be the key for the next-generation medical treatments.

\subsection{Tissue properties verification, ``prediction'' and control}
Now we discuss the self-adaptiveness of biological systems, particularly focusing on the bone remodeling process. In self-adaptive systems the next behavior of a component is determined by feedback information on the environment in which the component itself acts. We show that many biological phenomena intrinsically exhibit a self-adaptive nature, which is closely related to the concepts of regulation and shape. 

Despite bone remodeling occurs asynchronously at various sites, bone shapes and size are strictly controlled, suggesting that several multiscale controls exist. Therefore bone remodeling is a multi-level process, where macroscopic (tissue) and microscopic (cellular, molecular) scales are closely inter-dependent. 
Generally speaking, in animal biology, the concept of shape is inherently linked to the existence of control mechanisms and therefore to self-adaptiveness. Tissue homeostasis derives from a regulation of the relative activities of bone cells, in particular osteoblasts and osteoclasts, which control bone deposition and resorption, respectively. Bone homeostasis is also regulated through the environment, could be the internal one, for example hormones (for example parathyroid hormone and vitamin D, which are the principal modulators in calcium homeostasis) or the external one, for example external mechanical forces (for example physical activity, radiation such as the case of cancer patients, different gravity such as the case of astronauts~\footnote{see http://www.sciencedaily.com/releases/2006/07/060717091230.htm}). 

For many aspects the abstraction of cells, or any autonomous biological entity, could be {\em agents}. Following this, we take into account also the {\em environment} where cells, as agents, are acting and interacting. In turn, the environment could be seen as the entity that interacts with the systems (e.g. molecular, cell, tissue, organ, etc.) incorporated. We can generalize further if we see the environment in terms of rules that drive the emerging behavior of the whole systems, e.g. the parameters space that affects the shape of the whole system. We propose a self-adaptive system as a pair of {\em $<$environment, agent$>$}, the environment is represented by the parameter space and the functional constraints of the whole system (e.g. tissue), while the agent (e.g. a cell or set of cell) represents the system that during its dynamics, changes the parameters values of the environment and thus affects the satisfaction of the constraints. Generally speaking, an {\em agent} is a system that is located in some {\em environments}; is capable to {\em perceive} environmental information through his sensors; according to such information it {\em selects} the best action; and finally it {\em acts}, through his actuators, on the environment in order to meet its design objectives.
\begin{figure}[ht]
  \begin{center}
  \includegraphics[angle=0,width=0.7\textwidth]{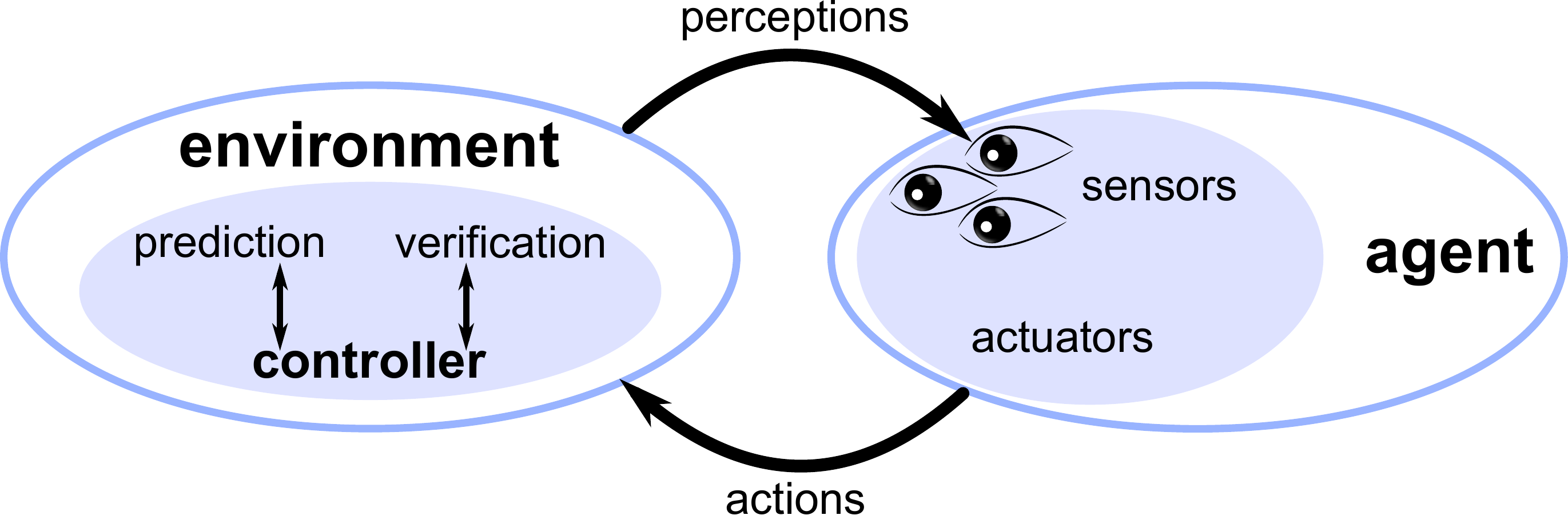}
  \end{center}
  \caption{Graphical representation of a self-adaptive system}
  \label{img:self-adaptive}
\end{figure}
With reference to Figure~\ref{img:self-adaptive} there are several concepts that form a correspondence between the system with self-adaptiveness and the bone system. First of all, shape maintenance is related to both the species evolution and the organism development, which is termed the evo-devo. The evolutionary basis provides a sort of bone strength ``prediction'', encoded in the genome while the development basis represent a sort of verification related to the conditions of the organism during his life. 
In the bone remodeling process, the sensing activities which trigger the control system have also a multiscale nature. In particular, the osteocytes in the bone matrix sense micro-fractures due to the apoptosis of the cells close to it and to the disruption of the canaliculi network.
The regulative mechanisms operate in a dynamical manner due to the BR process. Abnormalities in the bone size and shape are therefore related to important pathologies. Another important biological principle tells us that the structural information (i.e. the shape, size and orientation) drives the function. This link between the shape and the function in biology defines the self-adaptive nature of bone remodeling. Hence, the verification of properties on functional disruption could be used to assess disease severity.

Self-adaptiveness is a common characteristic of biological systems, since it implies the capability of the involved entities to react properly with respect to the environment in which they are located, and it implies regulative mechanisms able to ensure critical functional properties (e.g. tissue homeostasis). Moreover many biological principles such as \textit{morphogenesis} state that an organism develops its own shape according to the function, the role it must accomplish. For these reasons, we believe that a shape-based formal verification may give important insights into many unanswered questions of developmental biology. 

\subsection{Potentialities for translational medicine modeling}
We stress that multiscale and self-adaptive properties found in biological processes should inspire analogous properties in formal methods. The modeling of osteoporosis as a bone remodeling disruption is a prototype translational medicine problem which has offered us several points to discuss how formal methods should develop in order to provide better answer to the growing field computational medicine. 

The multi-level approach allows to manage simultaneously the concept of environmental parameters (at tissue level) and the concept of agents that interact within an environments (at cellular level). The cellular interactions, driven by the environmental parameters give as result a change in the parameters themselves that in turn drive new interactions.

The multi-level analysis, shifted from global to local and vice-versa, helps in monitoring the emerging of diseases such as the osteoporosis. In the local case, the probabilistic and temporal parameters should be present in the syntax of the modal operators and the interpretation of the resulting formulas should be qualitative, in the sense that it should return a truth value, e.g. balance between osteoclasts and osteoblasts. By contrast, in the global case no probabilistic and temporal parameters should be present in the syntax but the interpretation of the usual formulas should be quantitative, in the sense that it should return a number that measures how much a formula is satisfied, e.g. the bone mineral density.  The \textit{quantitative analysis} is manly based on the bone density, until under a certain threshold no osteoporosis, while over alert.  In the case of no alert, to be sure of the healthy state, a low-level \textit{qualitative analysis}, at cellular level, should be employed to verify the balance between osteoclast and osteoblast populations.
\\ \ \\ \textbf{Acknowledgment}\\
The authors thank Marco Viceconti for stimulating and invaluable discussions on bone remodelling accomplished at Istituti Ortopedici Rizzoli in Bologna. Nicola Paoletti thanks RECOGNITION: \textit{Relevance and cognition for self-awareness in a content-centric Internet (257756)}, funded by the European Commission within the 7$^{th}$ Framework Programme (FP7); and the EC-funded HPC-Europa2 programme, for supporting his visit to the Edinburgh Parallel Computing Centre (EPCC) and to the Computer Laboratory at the University of Cambridge.
\bibliographystyle{eptcs}
\bibliography{compmod}
\end{document}